\documentclass[runningheads]{llncs}
\usepackage{graphicx}
\begin{document}
\title{Failure type detection and predictive maintenance for the next generation of imaging atmospheric Cherenkov telescopes\thanks{Supported by "Istituto Nazionale di Astrofisica" (INAF) Mini Grant.}}
\titlerunning{FTD and PdM for the next generation of IACTs}
%
\author{Federico Incardona\inst{}\orcidID{0000-0002-2568-0917} \and
Alessandro Costa\inst{}\orcidID{0000-0003-0344-8911} \and
Kevin Munari\inst{}\orcidID{0000-0002-7062-8683}
}
\authorrunning{F. Incardona et al.}
%
\institute{INAF, Osservatorio Astrofisico di Catania, Via S Sofia 78, I-95123 Catania, ITALY
 \email{federico.incardona@inaf.it} \\
}
\maketitle              
\begin{abstract}
The next generation of imaging atmospheric Cherenkov telescopes will be composed of hundreds of telescopes working together to attempt to unveil some fundamental physics of the high-energy Universe. Along with the scientific data, a large volume of housekeeping and auxiliary data coming from weather stations, instrumental sensors, logging files, etc., will be collected as well. Driven by supervised and reinforcement learning algorithms, such data can be exploited for applying predictive maintenance and failure type detection to these astrophysical facilities. In this paper, we present the project aiming to trigger the development of a model that will be able to predict, just in time, forthcoming component failures along with their kind and severity.

\end{abstract}
\section{Introduction}
When entering the atmosphere, the very-high-energy gamma rays generate cascades of secondary charged particles that, in turn, produce Cherenkov radiation that hits the ground on a wide area, in the order of many hundreds of square meters. To collect this radiation in the widest energy range and at the highest sensitivity, many telescopes of different sizes must be deployed. The next generation of imaging atmospheric Cherenkov telescopes (IACTs), whose main representative is the Cherenkov Telescope Array (CTA)~\cite{ref_science}, will be composed of hundreds of telescopes working together. 

Along with the scientific data collected by the telescopes, a large volume of housekeeping and auxiliary data coming from weather stations, instrumental sensors, log messages, LiDARs (light detection and ranging), FRAMs (photometric robotic atmospheric monitor), etc., will be collected as well. The multitude of sensors spread all over the instrumentation will make the next generation of IACTs some Internet of Things (IoT) environments. The complex software architecture required to ingest and consume the whole amount of data is currently being developed by exploiting cutting-edge technologies in the field of IoT and big data~\cite{ref_monaas}.

Sensor data are usually exploited for addressing systematic errors in scientific measures. However, driven by proper machine learning algorithms, they can be used for applying some innovative techniques of maintenance already used in Industry 4.0. 

\subsection{Predictive maintenance and failure type detection}
\textit{Predictive maintenance} (PdM) and \textit{failure type detection} (FTD) are data-driven approaches to maintenance that improve the life cycle of a system by predicting potential equipment malfunction and reducing its downtime. The goal of PdM is to enable the replacement of the system components \textit{just in time}, namely when they are close to failure. A further gain is achieved through FTD, which aims to predict not only the time but also the \textit{nature} of the fault of a system component, or rather, the kind of anomaly. 

\subsubsection{The project} aims to trigger the development of a \textit{prognosis and analysis} model, respectively for PdM and FTD, which will help the maintainers of the next generation of IACTs to lower the telescopes' inactivity to only a few percent in favor of an increased observational time. The model will be able to analyze the current state of the real-world system and describe its behavior across its lifetime for the prognosis of a future state. It will be based upon statistical and learning techniques, driven by time-series data coming from the \textit{condition-based monitoring} (CBM) system, the complex network of sensors committed to monitoring critical parameters such as vibrations, acoustic and infrared emissions, ultrasounds, temperatures, currents, torques, lubricant, and oil conditions, etc. In many cases, as in the one of CTA, the CBM system is included in the maintainability plan of the facilities, but the corresponding PdM and FTD applications are not conceived in the design of the monitoring software. Therefore, the proposed model is one of a kind.

\section{Project description}
The model will take advantage of a \textit{supervised machine learning} algorithm, in which the training set will be properly labeled. Given the complexity and variety of data, it will be based upon gradient-boosted decision trees. It will also be able to improve itself thanks to the real-time stream of data from the telescope by exploiting \textit{reinforcement learning}.

\subsection{Fault database}
In order to build, tune, test, and validate the model, clips of historical data recording component failures and anomalies (e.g., spikes in the supply current, mechanical overheating, etc.) must be gathered as much as possible, along with the system fault coverage (i.e., the percentage of every type of fault that has occurred during the system lifetime). The collected data must be then properly labeled. The goal of labeling is to obtain a \textit{fault database} (FDB) containing data that enables the machine learning algorithms to be trained properly and to make accurate predictions. A \textit{prediction window} must be identified in the time series, defining the time prior to the system fault within which the maintainer wants to be alerted. This is typically in the order of weeks, and it has to trade off the time needed to schedule a repair and the risk of replacing a component while it is still functional.

\subsection{Fault tree analysis}
A \textit{fault tree analysis} (FTA) has been already performed~\cite{ref_gambadoro_marocco,ref_gambadoro_ML4ASTRO}, as a preliminary task, to study the role of each component in causing potential system failure. The FTA is a schematic view of the whole system in which a possible top-level fault could result from a Boolean combination of a series of low-level events (not necessarily faults). Usually, the most critical components are three-phase motors, encoders, bearings, and pneumatic breaks. Although not part of the telescope per se, ancillary or infrastructural systems, such as weather stations, power plants, LiDARs, and FRAMs, are critical to the attainment of the scientific objectives. Also crucial is the central equipment, which allows the whole array to perform as a single instrument. 

An interesting outcome of the project is the possibility to widen the field of application of the model to other telescope facilities that employ the same list of critical components, or a subset of them. The aforementioned list of critical elements, indeed, is not specific to IACTs. Radio telescope arrays, e.g., the Square Kilometre Array (SKA)~\cite{ref_SKA}, are built in a similar fashion, and the model described so far could be easily tailored to them. 

%
%
%
%

\end{document}